# Mapping the Trap-State Landscape in 2D Metal-Halide Perovskites using Transient Photoluminescence Microscopy

*Michael Seitz, Marc Meléndez, Nerea Alcázar-Cano, Daniel N. Congreve, Rafael Delgado-Buscalioni, and Ferry Prins\**

M. Seitz, Dr. F. Prins
Condensed Matter Physics Center (IFIMAC) and Department of Condensed Matter Physics, Autonomous University of Madrid, 28049 Madrid, Spain
E-mail: ferry.prins@uam.es

Dr. M. Meléndez, N. Alcázar-Cano, Prof. R. Delgado-Buscalioni
Condensed Matter Physics Center (IFIMAC) and Department of Theoretical Condensed Matter Physics, Autonomous University of Madrid, 28049 Madrid, Spain

Dr. D. N. Congreve
Rowland Institute at Harvard University, Cambridge, Massachusetts 02142, United States
Electrical Engineering Department, Stanford University, Stanford, CA 94305, United States

Keywords: 2D, metal-halide, perovskites, exciton, trap-states, transport, diffusion

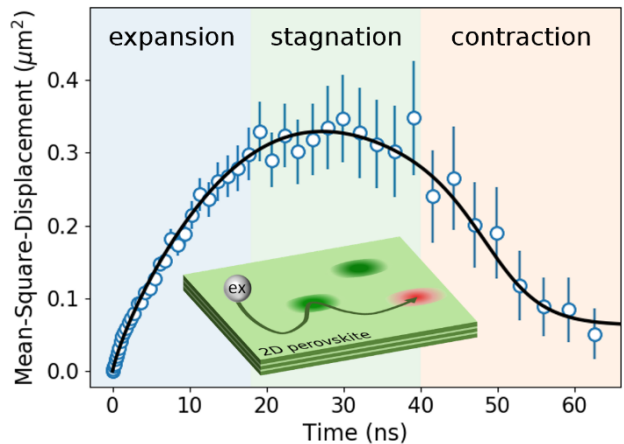

Transient microscopy is of vital importance in understanding the dynamics of optical excited states in optoelectronic materials, as it allows for a direct visualization of the movement of energy carriers in space and time. Important information on the influence of trap-states can be obtained using this technique, typically observed as a slow-down of the energy transport as carriers are trapped at defect sites. To date, however, studies of the trap-state dynamics have been mostly limited to phenomenological descriptions of the early time-dynamics. In this report, we show how long-acquisition-time transient photoluminescence microscopy can be used to provide a detailed map of the trap-state landscape in 2D perovskites, in particular when used in combination with transient spectroscopy. We reveal anomalous spatial dynamics of excitons in 2D perovskites, which cannot be explained with existing models for trap limited exciton transport that only account for a single trap type. Instead, using a continuous diffusion model and performing Brownian dynamics simulations, we show that this behavior can be explained by accounting for a distinct distribution of traps in this material. Our results highlight the value of transient microscopy as a complementary tool to more common transient spectroscopy techniques in the characterization of the excited state dynamics in semiconductors.

1. Introduction

Two-dimensional (2D) metal-halide perovskites have attracted significant attention as a more stable and versatile analog to their three-dimensional (3D) bulk counterparts.[1–4] Efficient solar cells with improved stability have been made from 2D/3D mixtures, as well as phase pure 2D perovskites with efficiencies above 21 and 18%, respectively.[5–8] Moreover, 2D perovskites are a promising material platform for light emitting applications,[9–12] as they offer high photoluminescence quantum yields, tunable and spectrally pure colors,[13,14] and solution processability.[15] 2D perovskites are described by their general chemical formula $L_2[ABX_3]_{n-1}BX_4$, where L is a long organic spacer molecule, A is a small cation (e.g. $Cs^+$, methylammonium, formamidinium), B is a divalent metal cation (e.g. lead, tin), X is a halide anion (chloride, bromide, iodide), and n is the number of inorganic octahedra that make up the thickness of the inorganic layer. The confinement of charge carriers into few-atom thick inorganic layers in 2D perovskites leads to strong quantum and dielectric confinement effects.[16] As a result, the optoelectronic properties of 2D perovskites are dominated by bound electron-hole pairs, called excitons.[17,18]

Understanding the spatial dynamics of the excitonic excited state is crucial in the optimization of device performance. For example, long diffusion lengths are required for solar harvesting as excitons need to reach charge separation sites, while short diffusion lengths are preferable for light-emitting technologies as they reduce the chances of encountering other excitons or non-radiative trapping sites.[19] Earlier this year, the first studies on exciton transport in these materials were reported using transient microscopy, directly visualizing the spatial displacement of the excitons as a function of time.[20–23] Deng et al. determined the diffusion constants of a set of 2D perovskites using transient absorption microscopy (TAM), showing how exciton transport improves for increasing inorganic layer thickness n.[20] In parallel, our own group reported on exciton transport in 2D perovskites with varying organic spacers and thicknesses using transient photoluminescence microscopy (TPLM), showing that exciton-phonon coupling is a key parameter in determining the intrinsic exciton transport in these materials.[21] Crucially, for bright materials like 2D perovskites, transient photoluminescence microscopy benefits from better signal to noise ratios as compared to, for example, transient absorption microscopy.[24] As a result, we were able to follow exciton transport over several nanoseconds after excitation and in this way extract crucial information about the transition from purely diffusive to trap-state limited transport.[21] This highlights the value of TPLM as a characterization tool for trap-states.

The ability to extract information on the trap-state dynamics from TPLM measurements was first reported by Akselrod et al., who observed a transition from normal diffusion to a regime of trap-state limited diffusion in organic single crystals.[25] They described the deacceleration using a classical one-dimensional diffusion model taking MSD(t) = $2D_0 t^\alpha$, with MSD being the mean-square-displacement of excitons, $D_0$ the diffusion coefficient, and $\alpha$ the diffusion exponent. For normal diffusion through a random walk one finds $\alpha = 1$ and a constant diffusivity ($D(t) = \frac{1}{2}\frac{dMSD(t)}{dt} = D_0$). For a trap-state limited regime, an $\alpha < 1$ accounts for a diffusivity that slows down with time ($D(t) = D_0 t^{\alpha-1}$), representing a subdiffusive behavior. Unfortunately, this phenomenological model does not allow any quantitative characterization of the trap-state properties. More recently, our group[21] and Folie et al.[26] independently introduced a model that allows to describe the decrease in diffusivity in terms of a trap density $1/\lambda^2$:

$$\text{MSD}(t) = 2\lambda^2 \left(1 - e^{-\frac{D_0}{\lambda^2}t}\right) \qquad (1)$$

This model allowed us to describe exciton diffusion in 2D metal-halide perovskites for times up to 8 ns. Importantly though, this model assumes deep trapping sites that do not allow for detrapping of excitons. Shortly after, Delor et al. reported an empirical model which accounts for traps that allow detrapping of excitons:[27]

$$\text{MSD}(t) = 2\tau_{turnover}(D_0 - D_{trapped})\left[1 - e^{-\frac{t}{\tau_{turnover}}}\right] + 2D_{trapped}t \qquad (2)$$

where $D_0$ is the diffusivity of the free excitons, $D_{trapped}$ is an average diffusivity that accounts for excitons being stuck at a trapping site for a certain time, and $\tau_{turnover}$ ($=\frac{\lambda^2}{D_0}$) is the characteristic trapping time. For $D_{trapped} = 0$ the model with deep trapping sites that do not allow detrapping is recovered (Equation 1).[21,26] Importantly, Equation 2 allows for the extraction of both a trap density $1/\lambda^2$ and the depth of the trap through $\Delta E_{trap} = \ln(D_0/D_{trapped}) \cdot k_B T$.[27]

In this report, we present transient photoluminescence microscopy (TPLM) measurements of exciton diffusion in $(PEA)_2PbI_4$ 2D perovskites across extended time scales and show that existing models cannot account for the observed complex spatial dynamics at later times. While the early time dynamics is characterized by a rapid expansion and the previously observed transition to a trap-state limited regime,[21] at later times, a full stagnation and even contraction of the MSD is observed. By combining our TPLM with transient spectroscopy and a continuous diffusion model, we show that the dynamics can only be explained by going beyond the conventional single-trap picture and accounting for a distinct distribution of trapping sites. This study highlights how the complementary use of spatial, spectral, and temporal information allows a more complete picture of the complex trap-state dynamics in 2D perovskites, reinforcing the position of transient microscopy as an increasingly important characterization tool for optoelectronic materials.

## 2. Results and Discussion

All measurements are performed on single crystalline flakes of phenethylammonium lead iodide $(PEA)_2PbI_4$. We grow large (~10-100s of micrometers) single crystals of $(PEA)_2PbI_4$ from supersaturated precursor solution under ambient conditions (see Methods section for details).[28,29] The single-crystals are then exfoliated and transferred to a microscope cover slip before each measurement to produce a freshly cleaved surface with minimal exposure to air. The cover slip on one side and the bulk of the crystal on the other side provide a form of self-passivation from oxygen and moisture during the measurements.[21]

TPLM is performed using a near-diffraction limited pulsed excitation laser ($\lambda_{ex}$ = 405 nm) to create a narrow initial exciton population. The image of the emission is projected outside the microscope with a total magnification of 330x. Spatially resolved photoluminescence lifetime measurements are then performed using an avalanche photodiode (APD) on a linear scanning stage. During measurements, the laser fluence is kept low (~250 nJ/cm$^2$) to prevent second order effects (e.g. Auger recombination) from influencing the observed dynamics.[21] The isotropic exciton transport properties of the inorganic plane allow us to measure a one-dimensional slice through the center of the exciton population and still capture the full exciton dynamics (see **Figure 1a**).[20]

**Figure 1b** shows the resulting normalized diffusion map, illustrating the time-dependent spatial distribution of the exciton population. We can quantify this behavior by fitting each time slice with a Voigt profile (see Methods for details) from which we extract the time evolution of the mean-square-displacement MSD(t) = σ(t)² – σ(0)² (see **Figure 1c**).[30] The slope of this curve is proportional to the diffusivity D, following $D(t) = \frac{1}{2}\frac{dMSD(t)}{dt}$. We observe an early regime (< 1 ns) of fast linear growth with a constant diffusivity D(t) = $D_0$, which is followed by a slower linear regime (~1-20 ns). Interestingly though, after around 20 ns we observe a full stagnation of the displacement and even contraction after around 40 ns. The early expansion regime is consistent with our earlier studies on 2D perovskites, in which exciton transport up to 8 ns was reported and described with Equation 1.[21] In the following, we will describe the shortcomings of the existing trap-state models and introduce a more rigurous one to account for our observations at times beyond 8 ns.

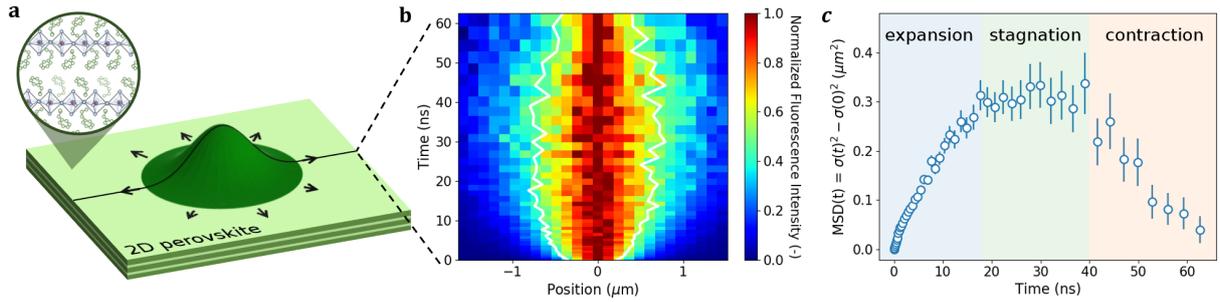

**Figure 1.** ((a) Illustration of transient photoluminescence microscopy (TPLM). A narrow exciton population is excited with a near-diffraction limited laser pulse in a (PEA)$_2$PbI$_4$ perovskite single-crystal (see inset for crystal structure). Recording the temporal evolution of the emission spot with a scanning avalanche photodiode allows to track both the spatial and temporal dynamics of the exciton population, which is proportional to the photoluminescence emission intensity at low fluences. The results of such a measurement is shown in b) Map of normalized emission intensity showing the broadening of the exciton population in space and time. The white line represents the full-width-half-max of the population. c) Mean-square-displacement (MSD) of the exciton population as a function of time. Initially the MSD increases due to excitons diffusing outwards. After around 20 ns the MSD stagnates and starts to decrease after around 40 ns.))

**Figure 2** shows the early time dynamics of the MSD, focusing on the expansion regime. The linear expansion at time < 1 ns is clearly visible, followed by a regime of a lower diffusivity, as excitons start getting trapped. In our previous work, we were able to successfully reproduce our experimental observations for early times (t < 8 ns) by assuming a simple model of deep traps that do not allow any detrapping of excitons (Equation 1, dashed line).[21] However, acquiring data for longer delay times, we see that our exciton diffusion data agrees better with a model that accounts for shallow traps, which do allow for detrapping of excitons (Equation 2, solid black line). This model captures both the normal diffusion ($D_0$ for t < 1 ns) and the transition to slower diffusion at later times (~1-20 ns) where excitons get temporarily stuck at trapping sites, which is represented by the average diffusivity $D_{trapped}$ (< $D_0$). For our $(PEA)_2PbI_4$ 2D perovskites we find $D_0$ = 0.223 cm$^2$/s, $D_{trapped}$ = 0.080 cm$^2$/s, and $\tau_{turnover}$ = 1.2 ns. The diffusivity $D_0$ for free excitons is consistent with our previously reported value of $(PEA)_2PbI_4$ synthesized with the same method.[21] Further, a Boltzmann ratio of 2.8 (= $D_0/D_{trapped}$) allows the estimation of a trap depth of $\Delta E_{trap}$ = 26 ± 2 meV (= ln(2.8)·$k_B$T).[27]

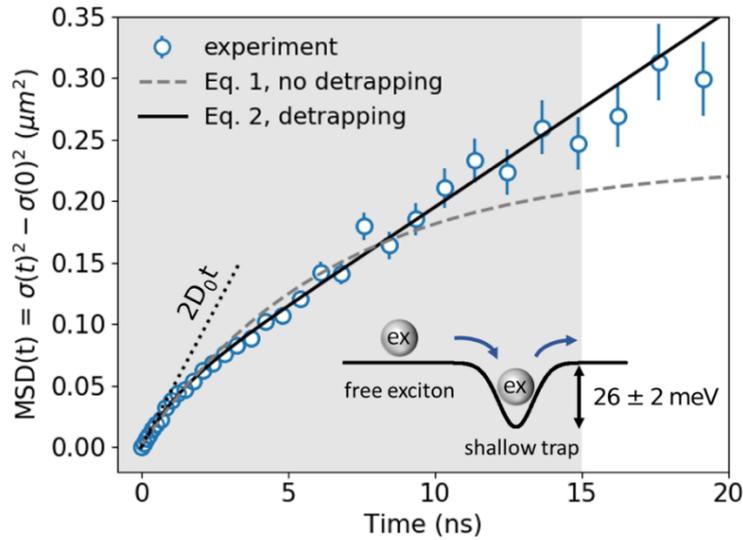

**Figure 2.** ((a) Mean-square-displacement (MSD) of the exciton population for t < 20 ns. Two linear regimes are observed: One at early times (t ≲ 1 ns) with fast diffusion of free excitons

and another one at later times (~1-20 ns), where the broadening is slower, due to excitons getting temporarily trapped in shallow trapping sites. Dashed line represents the model where excitons are not able to escape the traps (Equation 1; $D_0 = 0.177$ cm$^2$/s, $\tau_{turnover} = 6.5$ ns). Solid line accounts for detrapping of excitons (Equation 2; $D_0 = 0.223$ cm$^2$/s, $D_{trapped} = 0.080$ cm$^2$/s, $\tau_{turnover} = 1.2$ ns), which is a better representation of the experimental data. Fits were performed between 0 and 15 ns (shaded gray area). The inset shows a schematic of a free and a trapped exciton that can escape the trap. The ratio of $D_0/D_{trapped}$ and the Boltzmann relation can be used to estimate a trap depth of $\Delta E_{trap} = 26 \pm 2$ meV.[27]))

Interestingly, the trap-state energy that we obtain from transient microscopy compares favourably to the spectral shape of the photoluminescence spectrum. As shown in **Figure 3a**, the emission peak has a tail toward lower energy, which is often attributed to trap-state emission.[31–33] This tail is well reproduced using a fit with two Voigt functions with an energy difference of $23 \pm 2$ meV, close to the trap-depth of $\Delta E_{trap} = 26 \pm 2$ meV from the diffusion measurement.

To gain more insight into the spectral dynamics of the exciton population, we resolve the photoluminescence spectrum in time using a streak camera. As shown in **Figure 3b**, we observe a clear transient redshift of the emission ($\Delta E = 25 \pm 1$ meV), which can be interpreted as a shift in emission from free to trapped excitons. To confirm this, we fit each temporal slice to two Voigt functions, which allows us to extract the spectral weight of the two distributions at each point in time $t$ (**Figure 3c**). While the high-energy population displays prompt decay, the low-energy population shows an initial growth, consistent with an initially empty population of traps that are gradually being populated by diffusing free excitons. Importantly, we find excellent agreement between the time-scale of the transient red shift (double exponential fit in Figure 2b:

$\tau_1 = 0.9$ ns and $\tau_2 = 23$ ns) and the transition from normal diffusion to trap-state limited diffusion obtained from our TPLM measurements ($\tau_{turnover} = 1.2$ ns and $\tau_{stagnation} \approx 20$ ns). The good agreement of both the trap-state energies and the characteristic time scales in both transient microscopy and transient spectroscopy measurements suggests that the dynamics have the same physical origin and highlights that the two techniques can be used complementarily to obtain a more complete picture of the exciton dynamics in semiconductors.

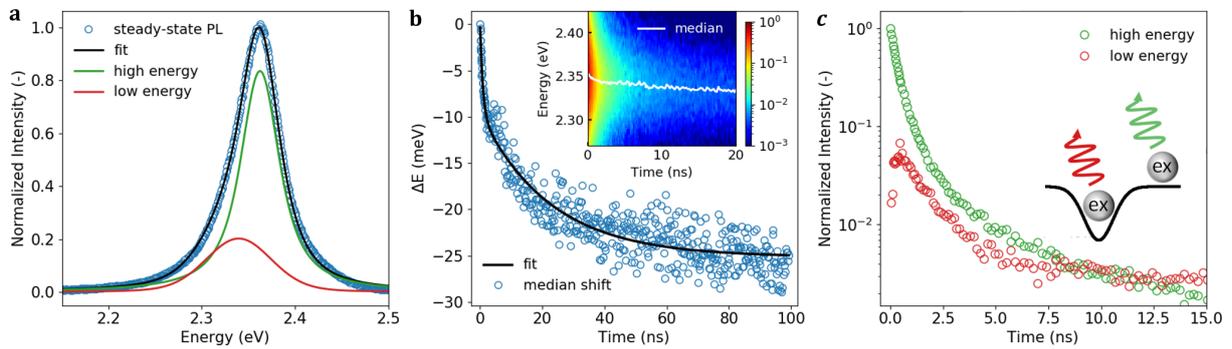

**Figure 3.** ((Photoluminescence spectroscopy. a) Steady-state photoluminescence of $(PEA)_2PbI_4$ and fit with two Voigt distributions. b) Median energy shift of the photoluminescence of $(PEA)_2PbI_4$ from a Streak camera image, which shows the photoluminescence as a function of energy and time (inset). The white line of the inset shows the median emission energy at each point in time. We fit the energy shift to a bi-exponential decay of $\Delta E = 25 \pm 1$ meV with decay constants of $\tau_1 = 0.9$ ns and $\tau_2 = 23$ ns. c) Spectral weight of the high and low-energy population as a function of time extracted from a fit of two Voight functions for each spectral slice in the streak camera image. Inset shows schematic of luminescent free and trapped excitons.))

While the shallow trap model of Equation 2 nicely reproduces the observed behavior for times up to 20 ns, it predicts a continuously increasing MSD(t) and therefore fails to capture our experimental observations at later time scales of stagnation (> 20 ns) and contraction (> 40 ns) of the MSD. To explain the observed stagnation and contraction regimes, we hypothesize that the MSD at later times is dominated by a sub-population of excitons that has been trapped for a prolonged amount of time. Longer trap-state lifetimes would correspond to deeper traps with slower detrapping rates. One important simplification of the shallow trap model (Equation 2) is the assumption of a single well-defined trap-state energy $\Delta E_{trap}$. In addition, it assumes that the radiative lifetime of the free and trapped excitons is identical. The transient spectroscopy results presented in Figure 3b show that while the red shift fits reasonably well to a double exponential decay, a slow gradual red-shift persists at longer times. This suggests that a distribution of trap-states around $\Delta E_{trap}$ is present, rather than a single trap with a fixed energy of $\Delta E_{trap}$. Moreover, from population dynamics presented in Figure 3c, trapped excitons appear to have a slower decay than the free exciton population.

To test the influence of a lower radiative decay rate for traps and a trap-state distribution on the MSD, we numerically solve the rate-equations and perform Brownian motion simulations (see Methods and SI for details). We assume a Gaussian distribution of trap-state energies that is centered around $\Delta E_{trap} = 25$ meV and has a width of $\sigma_{trap} = 35$ meV, corresponding to the trap-state emission that we extracted from the steady-state photoluminescence in Figure 3a. The detrapping rate is modelled as a thermally activated process using the Arrhenius equation: $\mu(\Delta E_{trap}) = A e^{-\frac{\Delta E_{trap}}{k_B T}}$ (see Figure S1 for the resulting rate distribution). We fit the experimental data using the trapping rate $\nu$, the proportionality constant A, and radiative decay rates $\gamma_{free}$ and $\gamma_{trap}$ as the fitting parameters (see **Figure 4a**). The result of the simulation is shown in **Figure 4b** along with the experimental data. Indeed, we find that we can reproduce the observed stagnation and decrease in MSD up to 40 ns, by accounting for a distribution of trap-state energies (black line with $D_0 = 0.223$ cm$^2$/s, $\nu = 1.4$ ns$^{-1}$, A = 12 ns$^{-1}$, $\gamma_{free} = 1.1$ ns$^{-1}$, $\gamma_{trap} = 0.7$ ns$^{-1}$), rather than of a single trap-state energy (orange line with $D_0 = 0.223$ cm$^2$/s, $\nu =$ 0.45 ns$^{-1}$, μ = 0.57 ns$^{-1}$, $\gamma_{free}$ = 1.1 ns$^{-1}$, $\gamma_{trap}$ = 0.7 ns$^{-1}$). We find that both a lower radiative decay rate of the traps ($\gamma_{trap}$ < $\gamma_{free}$) and a distribution of trapping states ($\Delta E_{trap}$ = 25 meV, $\sigma_{trap}$ = 35 meV) is necessary to successfully reproduce the later time dynamics of the MSD (see Figures S2,S3).

The simulations suggest that the stagnation and contraction of the MSD is a result of a trap-state distribution with different decay dynamics. At early times the MSD is dominated by quickly moving free excitons, that decay rather quickly and only get temporarily stuck in shallow traps, while the later times are dominated by excitons that are stuck in deeper traps that decay more slowly. We note that the final steep decay of the MSD observed in experiments (> 40 ns) is not fully captured by the model. Moreover, the photoluminescence decay of the model is faster than the experimentally observed values (Figure S4). This suggests that deeper traps beyond the Gaussian distribution or a distribution of radiative rates of the traps may need to be considered, both of which would lead to an emphasis of the contraction of the MSD. To illustrate this, we simulated a case where we introduced a small additional number of deep traps with an even slower radiative and minimized detrapping rate ($\gamma_{deep}$ = 0.5 ns$^{-1}$, $\mu_{deep}$ = 0 ns$^{-1}$). These traps do not impact the early time dynamics but do matter at later times as they are still present when most other excitons have already decayed, allowing to reproduce the steep decay of the MSD at later times (Figure 4b). The agreement between model and experiment is emphasized by the log-log representation of the MSD in Figure S5, showing an excellent fit over almost three orders of magnitude.

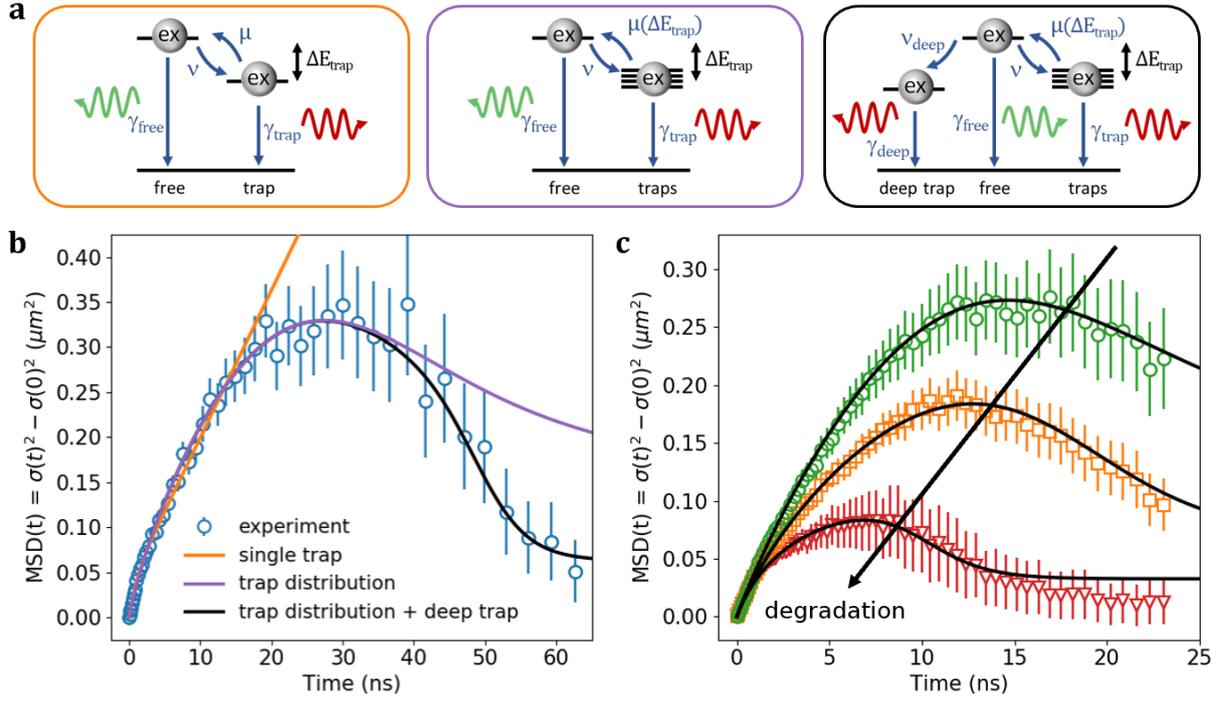

**Figure 4.** ((a) Illustration of the rates which are accounted for in the different models. b) Mean-square-displacement (MSD) of the exciton population as a function of time. Solid lines represent fits of models with different complexity. Orange line only accounts for a single trap-state with $D_0 = 0.223$ cm$^2$/s, $\nu = 0.45$ ns$^{-1}$, $\mu = 0.57$ ns$^{-1}$, $\gamma_{free} = 1.1$ ns$^{-1}$, $\gamma_{trap} = 0.7$ ns$^{-1}$. Purple line accounts for a trap distribution $D_0 = 0.223$ cm$^2$/s, $\nu = 1.4$ ns$^{-1}$, $A = 12$ ns$^{-1}$, $\gamma_{free} = 1.1$ ns$^{-1}$, $\gamma_{trap} = 0.7$ ns$^{-1}$, $\Delta E_{trap} = 25$ meV, $\sigma_{trap} = 35$ meV. Black line is the same as the purple line but includes a rare deep trap with $\nu_{deep} = 5 \cdot 10^{-8}$ ns$^{-1}$, $\gamma_{deep} = 0.5$ ns$^{-1}$. c) Three TPLM measurements on a different 2D perovskite flake, where the flake was sequentially degraded through prolonged exposure to high intensity blue light. Black lines represent fits with the model depicted in the right panel of panel a.))

Finally, we demonstrate how this model can be used to gain physical insight into the changes in the trap-state landscape upon degradation of the material. For this, we performed a series of TPLM measurements on a 2D perovskite flake that was sequentially photo-degraded between measurements using a high-intensity continuous wave blue LED (see **Figure 4c**). We find that the stagnation and contraction regimes are emphasized with increasing degradation, as excitons get more readily stuck at the additionally introduced trapping sites. Specifically, fitting the three curves with the model of a trap distribution including a rare deep trap (right panel of Figure 4a), we show that degradation leads to a higher trap-state density and introduces deeper traps, which is reflected in an increase of the trapping rates ($\nu$ and $\nu_{deep}$) and a broadening of the trap distribution ($\sigma_{trap}$), as well as a decrease in the radiative rate $\gamma_{deep}$ (see Table S1 for all fit parameters).

### 3. Conclusion

In summary, we have shown that TPLM is a powerful technique to study the exciton dynamics and material properties of semiconductors. Further, the good agreement of trap-state energies and characteristic time scales in both TPLM and transient spectroscopy suggests that the observed dynamics have the same physical origin and highlights that the two techniques can be used complementarily to gain a more complete picture of the exciton properties in semiconductors, with insights in both spatial and spectral dynamics. From our different measurements, we find that trap-states in $(PEA)_2PbI_4$ have an average depth of around $\Delta E_{trap}$ = 25 meV. TPLM revealed an anomalous stagnation and contraction regime of the MSD of exciton for t > 20 ns, which cannot be explained with previous descriptions of trap-state limited exciton transport. Using a continuous diffusion model and Brownian dynamics simulations we show that these dynamics can be explained by accounting for a trap distribution with a slower radiative decay than the free excitons.

### 4. Methods

*Sample preparation:* Chemicals were purchased from commercial suppliers and used as received. $(PEA)_2PbI_4$ single crystals were synthesized under ambient laboratory conditions following the over-saturation techniques reported previously.[28,29] In a nutshell, the precursor salts PEAI (Sigma Aldrich, 805904-25G), and $PbI_2$ (Sigma Aldrich, 900168-5G) were mixed in a stoichiometric ratio of 2:1 and dissolved in γ-butyrolactone (Sigma Aldrich, B103608-500G). The precursor solution was heated and kept at 70 °C for 2-3 days. The solution was cooled down to room temperature and drop cast on a glass slide, which was heated to 50 °C on a hotplate. After the solvent was evaporated, crystals with crystal sizes of up to several hundred microns were formed. The single crystals were mechanically exfoliated using the Scotch tape method and transferred to cover slips for the subsequent optical inspection in the microscope.

*Transient photoluminescence microscopy (TPLM):* Exciton diffusion measurements were performed following the same procedure as reported previously.[25,30] In short, a near-diffraction limited exciton population was created with a 405 nm pulsed laser (PicoQuant LDH-D-C-405, PDL 800-D) and a 100x oil immersion objective (Nikon CFI Plan Fluor, NA = 1.3) in a Ti2 inverted Nikon microscope. Photoluminescence of the exciton population was imaged on an avalanche photodiode (APD, Micro Photon Devices PDM, 20 µm x 20 µm) with a total magnification of 330x. As a result, the demagnified image of the APD had an effective area of around 60 x 60 nm (= 20 µm / 330), which was scanned through the middle of the exciton population in 120 nm steps, recording a time trace in every point. A Pico-Harp 300 timing board was used for the synchronization of the laser and APD for time correlated single photon counting. To minimize laser induced degradation of the perovskite crystals, the perovskite flake was scanned using an x-y-piezo stage (MCL Nano-BIOS 100), covering an area of 5 x 5 µm. TPLM were performed with a laser fluence of around 250 nJ/cm$^2$ and a 10 MHz laser repetition rate for all measurements, with the exception of the measurements presented in Figure 4c where we used 50 nJ/cm$^2$ and a 40 MHz. The time binning of the experimental setup was set to 4 ps and a nonlinear software binning was applied before analyzing the data. The bin size $n_k$ was calculated according to $n_k = \text{int}(0.3k^2 + 16) \to [16, 16, 17, 18, 20, 23, 26, 30, 35, …]$. The data was evaluated following the same procedure as previously reported by Seitz et al.[21]

*Transient photoluminescence spectroscopy:* Streak camera measurements were performed in a Ti2 inverted Nikon microscope. The sample was excited with a Hamamatsu pulsed laser (379 nm, 81 ps pulse width, 5 MHz, < 5 nJ/cm$^2$) and the emission spectrum was collected with a Hamamatsu C10627 streak unit, which was coupled with a C9300 digital camera, and a SP2150i spectrograph (Princeton Instruments). To obtain both high time resolution at early times and data for longer times we acquired two streak camera images. One with a time window of 20 ns and a time resolution of 42 ps (= 20 ns / 479 pixel) and one with 100 ns and a time resolution of 209 ps (= 100 ns / 479 pixel). We stitched the two measurements together at 8 ns. This allowed us to get data with both good time resolution at early times and data for long time delays (up to 100 ns). For the early time window we applied nonlinear software binning in time with the bin size increasing as $n_k$ = int(0.05k$^2$ + 1) → [1,1,1,1, 1,1,1,1,1,1,1,1,1,1, 2, 2, 2, 2, 2,…]. Fitting of the Streak camera image with two Voigt profiles $V(\mu, \Gamma, \sigma, c; t, A_k) = A_k \left( \frac{\frac{\Gamma}{2}}{(t-\mu)^2 + \frac{\Gamma^2}{4}} * e^{-\frac{(t-\mu)^2}{2\sigma^2}} \right) + c$, where μ the center of the distribution, Γ is the fwhm of the Lorentzian part, $\sigma^2$ is the variance of the Gaussian part, c is the offset c due to background noise, t is the time, and $A_k$ is the proportionality factor, which changes in time. μ, Γ, $\sigma^2$, and c were kept constant for all times for the two distributions.

*Continuous diffusion model (numerical integration of rate equations):* To describe the phenomenon in terms of rate equations, we discretize the escape rate distribution into a set of values $\mu_i$ (i = 1, 2, . . . , N) and track the concentration of free and trapped excitons in time and space. We use c(r, t) to represent the concentration of free excitons at time t and distance r from the initial laser pulse, and a different function $c_i$(r, t) for each possible escape rate $\mu_i$. Concentrations varied due to diffusion, trapping, escaping and decay through recombination. We integrated the equations numerically with a simple forward Euler time-stepping algorithm.

*Brownian motion simulations:* To model the dilute limit for excitons, we performed Brownian simulations of non-interacting particles diffusing through a landscape of traps. When an exciton reached a trap, it stopped for a random time determined by its escape rate, which was extracted from a log-normal distribution consistent with the Arrhenius relation between escape rate and trap energy, and a Gaussian distribution for the trap energies (with mean and standard deviation taken from the experiments). We represented free diffusion trajectories with the standard stochastic differential equation for Brownian motion in the Itô interpretation. The displacement of a free exciton during the time $\Delta t$ thus equals $\Delta \boldsymbol{r} = \sqrt{2D}d\boldsymbol{W}$, where D is the diffusion coefficient and $d\boldsymbol{W}$ is taken from a Wiener process, such that $\langle d\boldsymbol{W} d\boldsymbol{W}\rangle = \Delta t$. Numerical integration was carried out with the straightforward Euler-Maruyama method. The traps were distributed randomly and uniformly on a two-dimensional grid. In the dilute regime, once the traps are made small enough, the results become insensitive to the choice of trap size. The results of many trajectories were then averaged and the one-dimensional MSD (i.e. $\text{MSD}(t) = \frac{1}{2}\left(\text{MSD}_x(t) + \text{MSD}_y(t)\right)$) was compared to the experimental data.

## 5. Acknowledgements

This work has been supported by the Spanish Ministry of Economy and Competitiveness through the "María de Maeztu" Program for Units of Excellence in R&D (MDM-2014-0377). M. S. acknowledges the financial support through a Doc.Mobility Fellowship from the Swiss National Science Foundation (SNF) with the grant number 187676. In addition, M.S. acknowledges the financial support of a fellowship from "la Caixa" Foundation (ID 100010434). The fellowship code is LCF/BQ/IN17/11620040. M.S. has received funding from the European Union's Horizon 2020 research and innovation program under the Marie Skłodowska-Curie grant agreement No. 713673. F.P. acknowledges support from the Spanish Ministry for Science, Innovation, and Universities through the state program (PGC2018-097236-A-I00) and through the Ramón y Cajal program (RYC-2017-23253), as well as the Comunidad de Madrid Talent Program for Experienced Researchers (2016-T1/IND-1209). M.M., N.A., and R.D.B.

acknowledge support from the Spanish Ministry of Economy, Industry and Competitiveness through Grant FIS2017-86007-C3-1-P (AEI/FEDER, EU). D.N.C acknowledges the support of the Rowland Fellowship at the Rowland Institute at Harvard University.## 6. References

**2020**, *11*, 8565.

# Supporting Information

**Mapping the Trap-State Landscape in 2D Metal-Halide Perovskites using Transient Photoluminescence Microscopy**

*Michael Seitz, Marc Meléndez, Nerea Alcázar-Cano, Daniel N. Congreve, Rafael Delgado-Buscalioni, and Ferry Prins\**

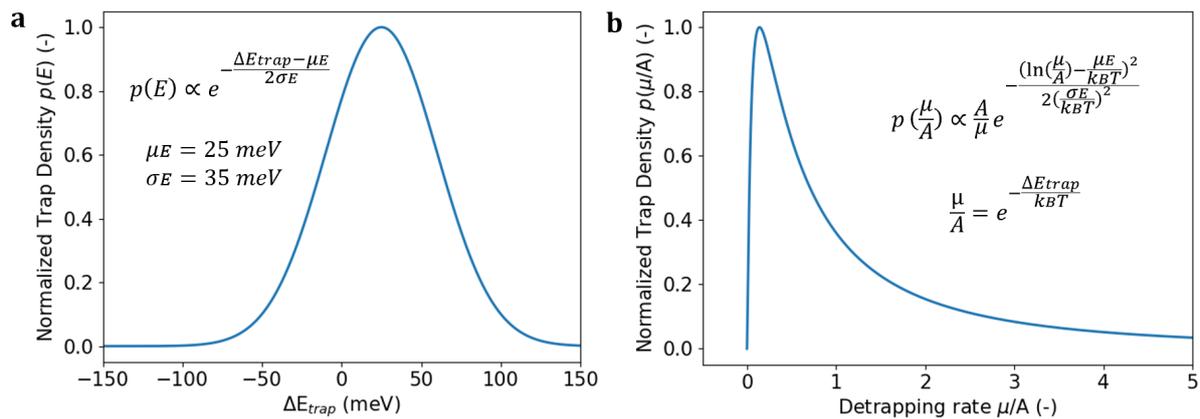

**Figure S1.** ((Trap-state population as a function of trap depth $\Delta E_{trap}$ (a) and detrapping rate µ (b).))

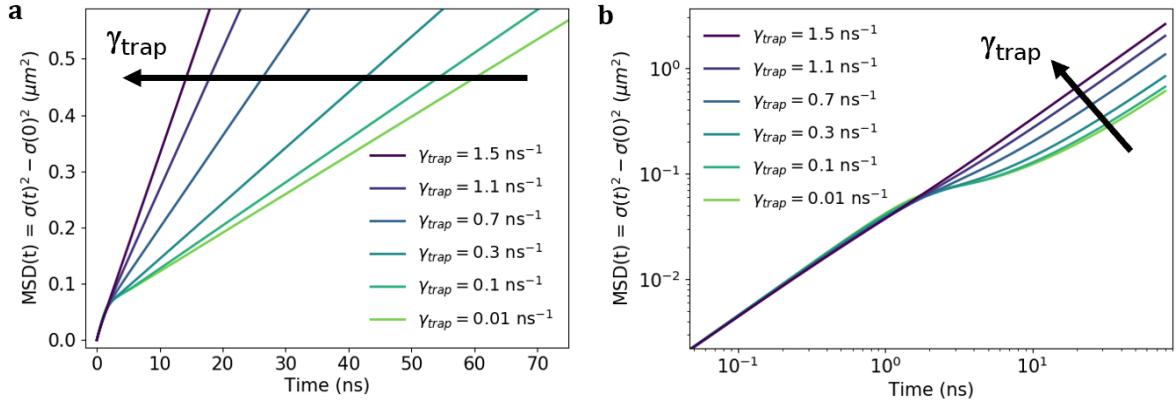

**Figure S2.** ((Simulation for diffusion with a single trap-state (orange line in Figure 4ab) with different radiative lifetimes for the traps and the same parameters as in Figure 4b: $D_0 = 0.223$ cm$^2$/s, $\nu = 0.45$ ns$^{-1}$, $\mu = 0.57$ ns$^{-1}$, $\gamma_{free} = 1.1$ ns$^{-1}$. The curves highlight that the MSD keeps increasing at late times when only accounting for a single type of trap instead of a trap distribution. a) Linear scale. b) Double logarithmic scale.))

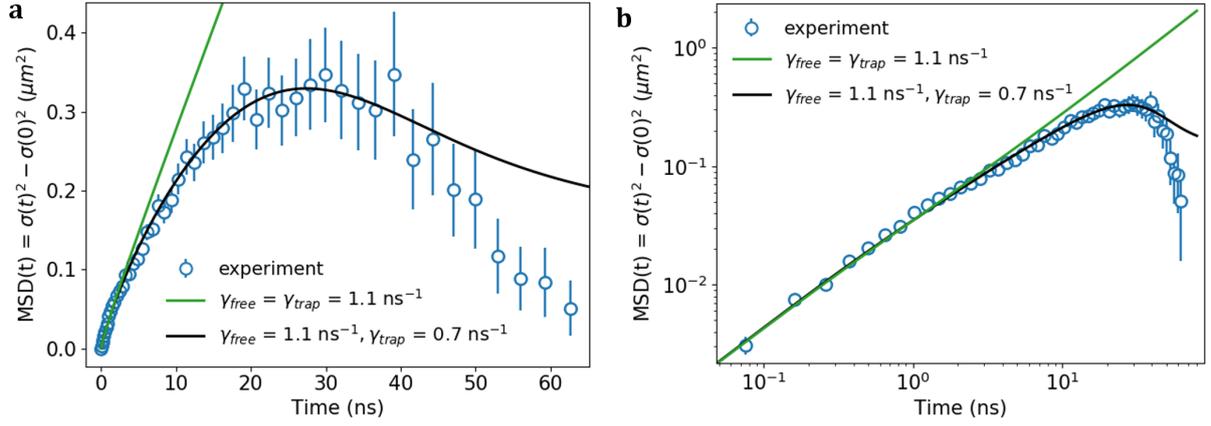

**Figure S3.** ((Mean-square-displacement (MSD) of the exciton population as a function of time. Open circles show the data from Figure 4b of the main text. Black line is the same as in Figure 4b, representing a model that accounts for both a trap-state distribution and a lower radiative decay rate for the traps $\gamma_{free} > \gamma_{trap}$. Green line represents curves that account for a distribution of trap-states for the detrapping rate but assumes $\gamma_{free} = \gamma_{trap} = 1.1$ ns$^{-1}$ for the radiative decay rates. Remaining parameters are the same as the black line here and in Figure 4b: $D_0 = 0.223$ cm$^2$/s, $\nu = 1.4$ ns$^{-1}$, $A = 12$ ns$^{-1}$, $\gamma_{free} = 1.1$ ns$^{-1}$, $\Delta E_{trap} = 25$ meV, $\sigma_{trap} = 35$ meV. The curves highlight that the MSD keeps increasing when not accounting for a decreased radiative decay rate of the trap-states. a) Linear scale. b) Double logarithmic scale.))

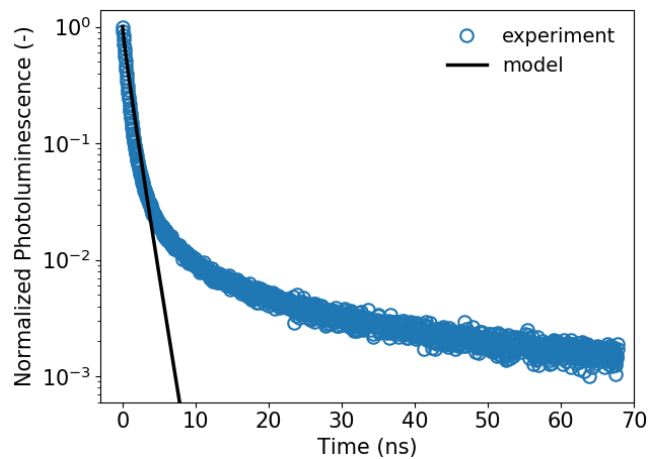

**Figure S4.** ((Photoluminescence decay of the experiments (markers) and the predicted decay by the model (solid line) for the same parameters as the purple line in Figure 4b: $D_0$ = 0.223 cm$^2$/s, $\nu$ = 1.4 ns$^{-1}$, A = 12 ns$^{-1}$, $\gamma_{free}$ = 1.1 ns$^{-1}$, $\gamma_{trap}$ = 0.7 ns$^{-1}$, $\Delta E_{trap}$ = 25 meV, and $\sigma_{trap}$ = 35 meV. Highlighting that traps with slower decay rates are needed to fully describe the late time dynamics.))

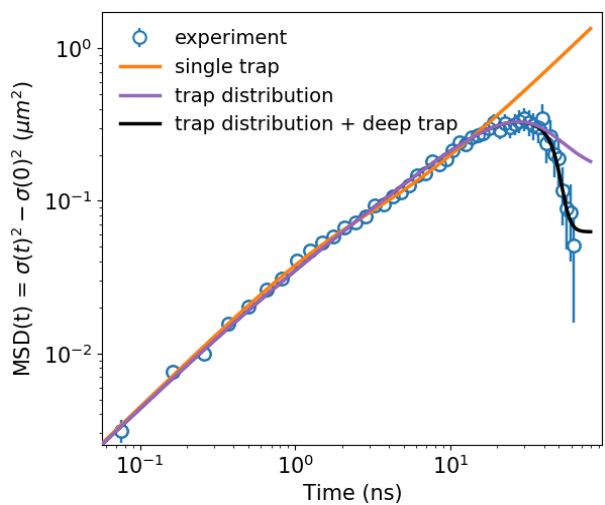

**Figure S5.** ((Log-log representation of Figure 4b showing the mean-square-displacement (MSD) of the exciton population as a function of time. Solid lines represent fits of models with different complexity.))

**Table S1.** ((Fits parameters of Figure 4c. Unchanged parameters are $D_0 = 0.223$ cm$^2$/s, $A = 12$ ns$^{-1}$, $\gamma_{free} = 1.1$ ns$^{-1}$, $\gamma_{trap} = 0.63$ ns$^{-1}$, $\Delta E_{trap} = 25$ meV.))

|  | $\nu$ (ns$^{-1}$) | $\sigma_{trap}$ (meV) | $\nu_{deep}$ (ns$^{-1}$) | $\gamma_{deep}$ (ns$^{-1}$) |
|---|---|---|---|---|
| circles | 0.55 | 35 | - | - |
| squares | 1.2 | 40 | 1e-3 | 0.5 |
| triangles | 1.6 | 60 | 1.5e-3 | 0.1 |

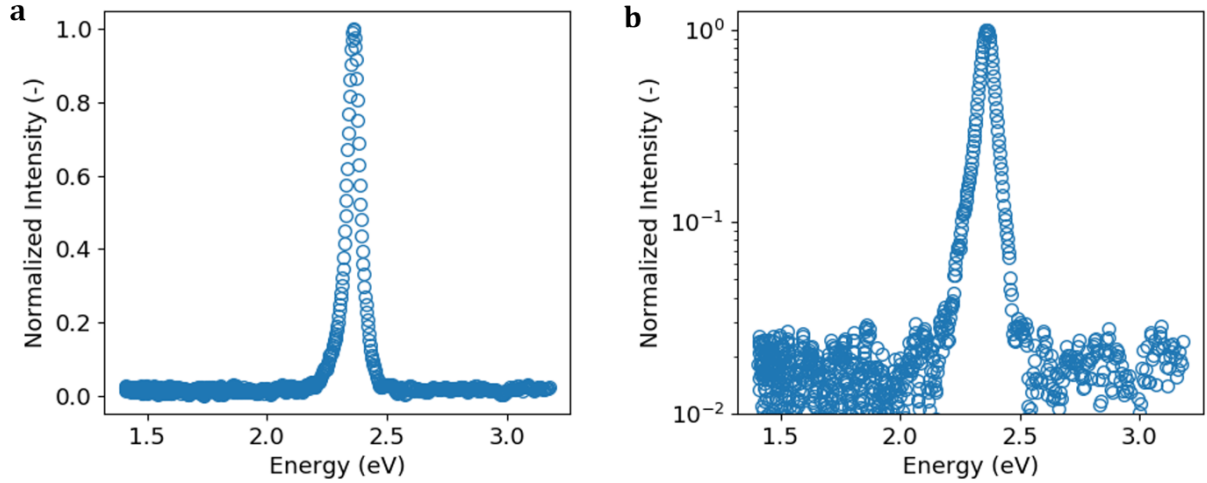

**Figure S6.** ((Steady state photoluminescence spectra on a linear (a) and semilogarithmic (b) scale, showing that all the emission is located between 2.2 and 2.5 eV.))

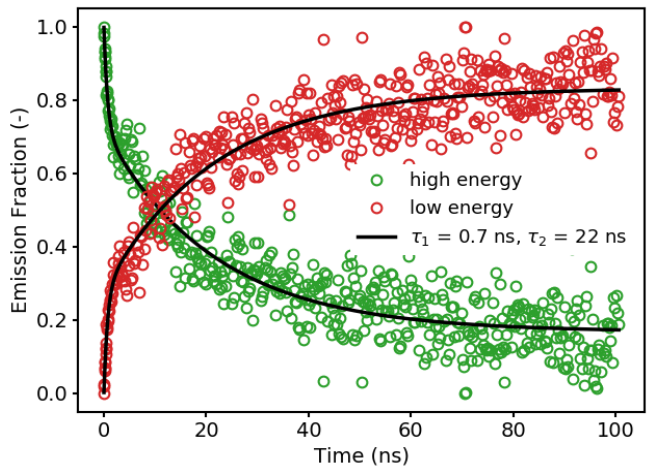

**Figure S7.** ((Emission fraction from the high and low energy population (Figure 3c) at each point in time obtained from the streak camera image in Figure 3b. Fit is a double exponential and reveals similar time dynamics to the median energy emission shift in Figure 3b $\tau_1 = 0.7$ ns and $\tau_2 = 22$ ns.))

**Continuous diffusion model (Numerical Integration of rate equations)**

A coarse-grained description of the process in terms of concentrations imagines each point in space large enough to contain many excitons and traps. Let c(r, t) represent the concentration of free excitons at distance r from the origin of coordinates and at time t. Assume that we begin with an initial concentration c(0, 0) = δ(0) of free excitons at the origin and describe their evolution. They will perform Brownian displacements through space with a diffusion constant $D_0$ and a fraction of the excitons will fall into traps at a rate given by ν.[21] We keep track of different types of trap according to their escape rates $\mu_i$, given by the log-normal distribution mentioned above (Figure S1b). Each type of trap has its own trapping rate $\nu_i$, determined from its probability ($\sum_i \nu_i = \nu$). Other excitons will decay through recombination, with rate $\gamma_{free}$. Furthermore, trapped excitons may escape with rate μi or decay with rate $\gamma_{trap}$. Therefore, the equations for the concentrations read

$$\frac{\partial c(r,t)}{\partial t} = \frac{D_0}{r}\frac{\partial}{\partial r} r \frac{\partial c(r,t)}{\partial r} - \left(\sum_i \nu_i + \gamma_{free}\right) c(r,t) + \sum_i \mu_i c_i(r,t),$$

$$\frac{\partial c_i(r,t)}{\partial t} = \nu_i c(r,t) - (\mu_i + \gamma_{trap}) c_i(r,t).$$

These equations were integrated numerically with a simple forward Euler time-stepping scheme, with space discretized into steps of $10^{-2}$ μm and time into steps of $10^{-3}$ ns. Concerning the discretization of escape-rate values, we observed that the long-time dynamics depended slightly on the resolution representing the smaller escape rates. Consequently, we used 900 different concentration functions $c_i$ to represent the lowest decile of escape rates and 100 for the remaining values. The black line in Figure 4b was obtained by adding in an extra concentration function for a very rare type of deep trap. Furthermore, it must be taken into account that, as in experiments, our results for MSD represent the exciton positions when the excitons radiatively decay by emitting a photon. In other words, we present the variance of photon emissions instead of exciton positions: $\text{MSD}_{\text{measured}}(t) = \frac{\int_0^\infty (\gamma_{free}c(r,t)+\sum_i \gamma_i c_i(r,t))\Delta t\, r^2\, rdr}{\int_0^\infty (\gamma_{free}c(r,t)+\sum_i \gamma_i c_i(r,t))\Delta t\, r\, dr}$, which is different from the actual MSD of the exciton distribution: $\text{MSD}_{\text{actual}}(t) = \frac{\int_0^\infty (c(r,t)+\sum_i c_i(r,t))\Delta t\, r^2\, rdr}{\int_0^\infty (c(r,t)+\sum_i c_i(r,t))\Delta t\, r\, dr}$. However, for only one exciton population or populations with the same radiative decay rates the two are equivalent ($\text{MSD}_{\text{experiment}}(t) = \text{MSD}_{\text{actual}}(t)$).

**Brownian dynamics simulations**

Brownian dynamics simulations have been performed to validate the predictions of our model or exciton diffusion in a plane having a uniform random distribution of traps. We simulate the stochastic differential equations for the Brownian motion with the following stochastic differential equation in the Itô interpretation: $\Delta r = \sqrt{2D}dW$, where $dW$ stands for the Wienner process, such that $\langle dW dW \rangle = \Delta t$. The diffusion coefficient $D$ corresponds to the diffusion coefficient of free excitons. Taking for granted that experiments were carried out in the dilute limit, we model the excitons in our simulations as non-interacting. Initially, an exciton diffuses freely until it reaches a trap, and there it stops. The probability of being trapped is given by the density of traps and the diffusion coefficient ($\lambda = D\, N_{traps}/L^2$, where $N_{traps}$ is the number of traps and $L$ is the system size). After a random time determined by the escape rate the exciton starts moving again. This detrapping rate comes from a log-normal distribution (Figure S1b) which is consistent with the Arrhenius relation between escape rate and trap energy ($\mu(\Delta E_{trap})$), and a Gaussian distribution for the trap energies (given by the experiments, Figure S1a). Also, an exciton can decay radiatively and its decay rate depends on whether the exciton is free or in a trap. The decay rates are constant and $\gamma_{free} > \gamma_{trap}$. The equation of motion has been numerically integrated by the well-known Euler-Maruyama algorithm and our results are not sensitive to the choice of trap size, since we are working in the dilute regime and the traps are small enough. Trajectories for many independent excitons have been computed and averaged to determine the MSD as a function of time. While the simulation of the MSD was done in two dimensions, we used the MSD in one dimension to match the experimental measurement: $MSD(t) = \frac{1}{2}\left(MSD_x(t) + MSD_y(t)\right)$. Futhermore, it must be taken into account that, as in experiments, our results for MSD represent the variance of exciton positions when they decay and emit photons. In other words, we present the variance of photon emissions instead of exciton positions.

The presence of a distribution of trap energies in our model makes good statistics very important in simulations. To improve the sampling, we need a very large number of excitons in our Brownian dynamics simulations and this implies a very long computation time. We compared the results to the theoretical model explained below using $10^7$ excitons. We were able to corroborate the theoretical predictions for the MSD up to 30 ns. From that moment on, too few excitons were left due to decay and the sampling was not representative.